\RequirePackage{silence}
\documentclass[aps,pra,twocolumn,showpacs,amsmath,amssymb,preprintnumbers,superscriptaddress,10pt,longbibliography]{revtex4-2}

\usepackage{graphicx}
\graphicspath{{figures/}}
\usepackage{SIunits}
\usepackage{hyphenat}
\usepackage[bookmarks=false]{hyperref}
\usepackage{color}
\usepackage[capitalise]{cleveref}
	\crefname{section}{Sec.}{Secs.}
	\Crefname{section}{Section}{Sections}
	\crefname{appendix}{Methods}{Methods}
	\Crefname{appendix}{Methods}{Methods}

\definecolor{eq_green}{RGB}{0,255,0}

\definecolor{pink}{RGB}{255,0,255}

\definecolor{darkgreen}{RGB}{0,170,0}

\definecolor{RED}{RGB}{255,0,0}

\newcommand{\tran}{^\top}

\begin{document}

\title{Intensity correlations in decoy-state BB84 quantum key distribution systems} 

\author{Daniil~Trefilov}
\email{dtrefilov@vqcc.uvigo.es}
\affiliation{Vigo Quantum Communication Center, University of Vigo, Vigo E-36310, Spain}
\affiliation{School of Telecommunication Engineering, Department of Signal Theory and Communications, University of Vigo, Vigo E-36310, Spain}
\affiliation{atlanTTic Research Center, University of Vigo, Vigo E-36310, Spain}
\affiliation{Russian Quantum Center, Skolkovo, Moscow 121205, Russia}
\affiliation{National Research University Higher School of Economics, Moscow 101000, Russia}

\author{Xoel~Sixto}
\affiliation{Vigo Quantum Communication Center, University of Vigo, Vigo E-36310, Spain}
\affiliation{School of Telecommunication Engineering, Department of Signal Theory and Communications, University of Vigo, Vigo E-36310, Spain}
\affiliation{atlanTTic Research Center, University of Vigo, Vigo E-36310, Spain}

\author{V{\'\i}ctor~Zapatero}
\affiliation{Vigo Quantum Communication Center, University of Vigo, Vigo E-36310, Spain}
\affiliation{School of Telecommunication Engineering, Department of Signal Theory and Communications, University of Vigo, Vigo E-36310, Spain}
\affiliation{atlanTTic Research Center, University of Vigo, Vigo E-36310, Spain}

\author{Anqi~Huang}
\affiliation{College of Computer Science and Technology, National University of Defense Technology, Changsha 410073, China}

\author{Marcos~Curty}
\affiliation{Vigo Quantum Communication Center, University of Vigo, Vigo E-36310, Spain}
\affiliation{School of Telecommunication Engineering, Department of Signal Theory and Communications, University of Vigo, Vigo E-36310, Spain}
\affiliation{atlanTTic Research Center, University of Vigo, Vigo E-36310, Spain}

\author{Vadim~Makarov}
\affiliation{Russian Quantum Center, Skolkovo, Moscow 121205, Russia}
\affiliation{Vigo Quantum Communication Center, University of Vigo, Vigo E-36310, Spain}
\affiliation{NTI Center for Quantum Communications, National University of Science and Technology MISiS, Moscow 119049, Russia}

\date{May 1, 2026}

\begin{abstract}
The decoy-state method is a prominent approach to enhance the performance of quantum key distribution (QKD) systems that operate with weak coherent laser sources. Due to the limited transmissivity of single photons in optical fiber, current experimental decoy-state QKD setups increase their secret key rate by raising the repetition rate of the transmitter. However, this usually leads to correlations between subsequent optical pulses. This phenomenon leaks information about the encoding settings, including the intensities of the generated signals, which invalidates a basic premise of decoy-state QKD. Here we characterize intensity correlations between the emitted optical pulses in two industrial prototypes of decoy-state BB84 QKD systems and show that they significantly reduce the asymptotic key rate. In contrast to what has been conjectured, we experimentally confirm that the impact of higher-order correlations on the intensity of the generated signals can be much higher than that of nearest-neighbour correlations.
\end{abstract}

\maketitle

\section{Introduction}
\label{sec:intro}

Quantum key distribution (QKD) represents a method for achieving information-theoretic security when sharing a confidential bit string, commonly referred to as a secret key, between distant parties \cite{bennett1984, xu2020, lo2014, pirandola2020}. Despite its theoretical security being rigorously proven \cite{lo1999, shor2000, koashi2009, renner2008}, practical implementations of QKD encounter challenges and limitations associated with current technology \cite{brassard2000, xu2020}, which might lead to security loopholes, or so-called side channels \cite{makarov2006, zhao2008, lydersen2010a, gerhardt2011, huang2019, huang2020, ruzhitskaya2021, ye2023}. To address these discrepancies between theory and practice, manufacturers of QKD equipment can apply improved security proofs that can handle device imperfections \cite{gottesman2004, lo2014, lim2014, tamaki2014, pereira2020, marquardt2023, curras-lorenzo2024, trushechkin2022, tupkary2024} and/or incorporate advanced hardware solutions \cite{dixon2017, ponosova2022, makarov2024}. Alternatively, the development and adoption of novel QKD protocols and methods, inherently resilient to specific vulnerabilities and quantum hacking attempts, offer another avenue. For example, measurement-device-independent (MDI) QKD closes all measurement loopholes without the need for theoretical characterization of the measurement unit \cite{lo2012}. Additionally, employing a twin-field (TF) QKD protocol has demonstrated the potential to significantly extend the achievable distance \cite{lucamarini2018, minder2019, wang2019, zhong2019, wang2022, liu2023}.

Nevertheless, despite these notable accomplishments, challenges remain to be addressed before QKD can attain widespread adoption as a technology \cite{xu2015a, diamanti2016, xu2020}. A crucial hurdle involves enhancing the secret key rate produced by existing experimental prototypes, a task affected significantly by the restricted transmissivity of single photons in optical fibers and the dead time of the receivers' detectors. For this reason, various experimental demonstrations have been conducted with an increased pulse repetition rate of the sources of several gigahertz \cite{grunenfelder2020, boaron2018}. Yet, within such a high-speed domain, the presence of memory effects in the optical modulators and their controlling electronics establishes correlations among the generated optical pulses, thus invalidating most security proofs. Significantly, this phenomenon introduces a security vulnerability in the form of information leakage. Fortunately, various security proofs have recently addressed the problem of pulse correlations \cite{pereira2020, zapatero2021, sixto2022, pereira2023, curras-lorenzo2023, pereira2024}, but they require a precise characterization of the source.

On the experimental side, a few recent works have quantified nearest-neighbour intensity correlations in various QKD system prototypes \cite{grunenfelder2020, yoshino2018, roberts2018} and showed that such correlations are, in general, not negligible. In contrast, higher-order correlations have been barely addressed in experimental settings \cite{kang2023, lu2023}, despite theoretical studies indicating that they can critically affect the system's security \cite{zapatero2021, sixto2022, pereira2024}. This gap is partly due to the commonplace assumption that intensity correlations decay rapidly with the correlation order. However, given their potential to significantly impact the effective secret key rate, more experimental efforts are needed to accurately assess and quantify higher-order intensity correlations in QKD systems, especially those already available on the market.

In this work, we experimentally study intersymbol intensity correlations in two industrial prototypes of decoy-state BB84 systems developed by two different vendors. We observe strong intensity correlations in both setups and apply a security proof that considers this imperfection \cite{sixto2022}. In doing so, we quantify the impact of this potential loophole on their performance in terms of secret key rate (SKR). Surprisingly, we find that in some cases higher-order correlations may affect the intensity of the emitted pulses more than nearest-neighbour correlations.

The paper is organized as follows. In \Cref{sec:experimental-setup}, we introduce the experimental setup we use to characterize the intensity correlations and describe the measurement procedure. There, we also explain the QKD protocol employed by the systems and define the assumptions we apply in the experiment. In \Cref{sec:results}, we present the experimental results revealing the intersymbol intensity correlations problem in both QKD systems studied. We then apply the security proof recently developed in \cite{sixto2022}, which takes into account this imperfection, and obtain asymptotic secret key rates in \cref{sec:theory}. We conclude in \cref{sec:conclusion}. The paper also includes a methods section with additional calculations.

\section{Experimental setup and measurements}
\label{sec:experimental-setup}

\begin{figure*}
\includegraphics{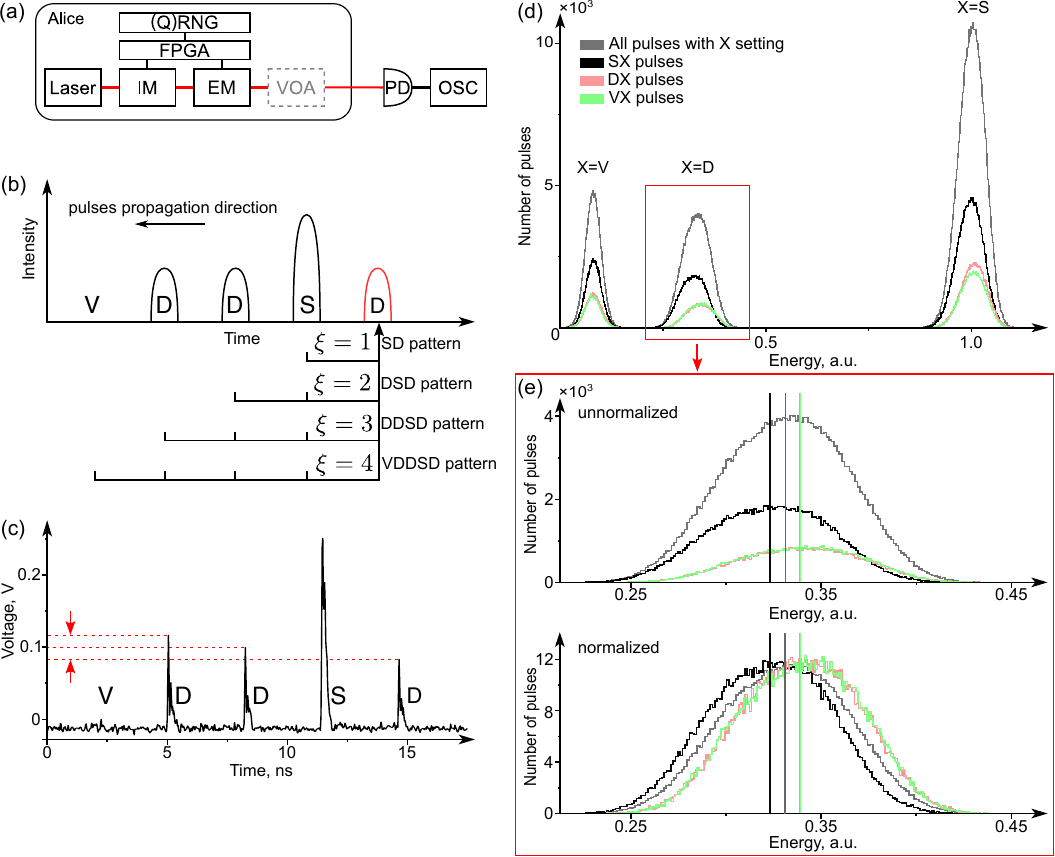} 
\caption{\label{fig:experimental-setup}Correlation measurement. (a)~Simplified scheme of the measurement setup for intensity correlations. Alice generates optical pulses with a laser diode (Laser). These pulses propagate through an intensity modulator (IM) and an encoding modulator (EM) and obtain a random intensity and encoding state according to the prescriptions of a decoy-state BB84 protocol. Both modulators are controlled with a field-programmable gate array (FPGA) that sets different operating voltage levels according to the signal received from a (quantum) random number generator (Q)RNG (one of the Alices tested is equipped with a classical pseudo-RNG). To ensure a classical energy level of the optical pulses at Alice's output, we remove a variable optical attenuator (VOA) from the optical scheme. The measurement setup consists of a fast photodetector (PD) and a digital high-bandwidth oscilloscope (OSC). (b)~Conceptual view of five consecutive optical pulses emitted by Alice. The intensity of the latest-emitted correlated pulse D depends on the previously emitted pulses, whose number determines the correlation length $\xi$. In this work, we shall consider correlations up to $\xi=6$. (c)~A short fragment of the recorded data oscillogram measured on system~B with five registered consecutive optical pulses. The intensity settings and the chronological order of the pulses are the same as in~(b). Dashed lines and arrows highlight the deviations in the maximum amplitude values of the D intensity setting pulses. (d)~Distributions of calculated energies for the studied intensity settings (gray) and $\xi=1$ in system~B. Each intensity setting, denoted by X, is represented by four distributions. We normalize the energy by the mean value of the S-state distribution. (e)~Zoomed-in fragment of~(d) with decoy-state distributions. We show both unnormalized (top) and normalized (bottom) groups of distributions. The latter group is normalized such that each distribution area's integral equals 1. The mean value for each distribution is marked by a vertical line of the same color. From these distributions, it follows that, in general, SD pulses have less energy than DD or VD pulses. The same trend can also be seen in~(c).}
\end{figure*}

We measure and analyze the intensities of modulated non-attenuated optical pulses produced by the source unit (Alice) of the two QKD systems considered. We shall refer to them as system~A and system~B. Although each of these systems is a complete engineering solution with its elaborately developed technical features, their crucial optical and electrical elements are similar. \Cref{fig:experimental-setup}(a) introduces one conceptual scheme that is accurate for the measurement sessions in both setups.

Both systems run a decoy-state BB84 protocol with three intensity settings \cite{wang2005a, lo2005, ma2005}. The applied intensity setting to the $k$-th pulse produced by system~A (system~B) is $a_{k_{A}}\in A_{A}=\{\mu_{A},\nu_{A},\omega_{A}\}$ ($a_{k_{B}}\in A_{B}=\{\mu_{B},\nu_{B},\omega_{B}\}$) with probability $p_{a_{k_{A}}}$ ($p_{a_{k_{B}}}$), where $p_{\mu_{A}}>p_{\nu_{A}}>p_{\omega_{A}}$ ($p_{\mu_{B}}>p_{\nu_{B}}\geq p_{\omega_{B}}$). The relation between the intensity levels in system~A (system~B) is $\mu_{A}>\nu_{A}>\omega_{A}\geq 0$ ($\mu_{B}>\nu_{B}>\omega_{B}\geq 0$), where $\mu$ represents the signal state (S), $\nu$ is the decoy state (D), and $\omega$ is the vacuum state (V). 

Performance speed restrictions and memory effects affect the core elements of Alice's setup that are involved in the preparation of the optical pulses she sends to Bob. These core elements include electro-optical modulators, high-speed electrical drivers, control motherboards, and CPUs. Due to these implementation limitations, an increase in the repetition rate of a QKD system can cause correlations between the emitted optical pulses. Hence, several parameters of an emitted optical pulse (i.e.,\ intensity, polarization, and phase) may depend on the parameters chosen to code the previously emitted pulses. In \cref{fig:experimental-setup}(b) we illustrate the concept of intensity correlations. We remark that, although this parameter is commonly labeled ``intensity'' in the literature on QKD, it actually represents the energy of the optical pulse. \Cref{fig:experimental-setup}(b) presents five consecutive optical pulses emitted by Alice with three intensity settings. In this model, the latest-emitted D pulse is correlated with the preceding pulses. We define the correlation length $\xi$ as the number of preceding consecutive pulses that condition the intensity of the considered pulse. In other words, if the correlation length of a QKD system is $\xi$, the pulse that is $\xi + 1$ positions away from the considered one, does not affect its intensity. In this context, we define higher-order intensity correlations as any correlation with $\xi \geq 2$. We also use the notion of \textit{pattern} to denote the combination of a pulse with its $\xi$ consecutive predecessors that affect its intensity. Assuming that the aforementioned D pulse in \cref{fig:experimental-setup}(b) exhibits correlations, different scenarios may arise depending on the value of $\xi$ in the system. The simplest case is first-order pulse correlations (i.e.,\ $\xi=1$), or so-called nearest-neighbour correlations, which correspond to the scenario where the intensity of a pulse depends on the intensity of the previous pulse [SD \textit{pattern} in \cref{fig:experimental-setup}(b)]. Similarly, the intensity of a pulse can be influenced by even earlier-emitted pulses along with its nearest neighbour. In the example provided by \cref{fig:experimental-setup}(b), this corresponds to second-order ($\xi=2$, \textit{pattern} DSD), third-order ($\xi=3$, \textit{pattern} DDSD), and fourth-order ($\xi=4$, \textit{pattern} VDDSD) correlations. As shown in this figure, we label each \textit{pattern} by listing the intensity settings of the preceding pulses in chronological order (from left to right), ending with the setting of the analyzed pulse itself (located in the rightmost position of the list). In all examples from \cref{fig:experimental-setup}(b), this final setting is D. While $\xi$ can in principle be arbitrarily large in a QKD system, in our work we limit the value of $\xi$ up to 4~(6) for system~A~(B), because the confidence intervals become too large for the higher values of $\xi$ in the measured data set. As we show below, the analyzed correlation \textit{patterns} for both systems exhibit noticeable deviations even at the highest considered values of $\xi$, indicating that the true value of this parameter in the tested systems, in principle, can be of an even higher degree. Nevertheless, our analysis can be straightforwardly adapted to any large value of $\xi$.

Following \cref{fig:experimental-setup}(a), in the measurement session, Alice of system~A (system~B) generates phase-randomized coherent pulses with a repetition rate of $40~\mega\hertz$ (several hundred $\mega\hertz$). Each of them is randomly modulated by an intensity modulator according to the prescriptions of the decoy-state BB84 protocol with three intensity settings. Then, the pulses pass through an encoding modulator (EM) and are randomly encoded in the BB84 states. To measure the intensity of the produced states, we connect Alice's output to a fast photodetector Picometrix PT-40A with DC to $38$-$\giga\hertz$ bandwidth (Thorlabs RXM40AF with $300$-$\kilo\hertz$ to $40\mbox{-}\giga\hertz$ bandwidth), which in turn is coupled to a digital oscilloscope Agilent DSOX93304Q with $33$-$\giga\hertz$ analog bandwidth and $80$-$\giga\hertz$ sampling rate (LeCroy SDA816Zi with $16$-$\giga\hertz$ analog bandwidth and $40$-$\giga\hertz$ sampling rate). The experimental data is recorded in the form of high-resolution voltage oscillograms with $12.5$ $(25)~\pico\second$ sampling period for system~A (system~B), containing hundreds of thousands of pulses. We show an example of the experimental data in \cref{fig:experimental-setup}(c). The shown data fragment consists of five consecutive optical pulses produced by system~B, which match the intensity settings presented in the concept scheme illustrated in \cref{fig:experimental-setup}(b). We mark the maximum amplitude value for each D pulse with dashed lines and highlight the difference with arrows. This difference hints that the intensities of the produced optical pulses are correlated.

We additionally process the raw experimental data and eliminate the instrument noise. While unfiltered noise does not influence the analysis of experimental results or affect our conclusions regarding the presence of correlation effects in the analyzed QKD systems as the mean values of the \textit{pattern} distributions remain essentially unchanged after filtering, it can bias the calculation of the secret key rate (SKR). This is because the SKR is highly sensitive to the width of the measured energy distributions. The filtering routines introduced below help reduce this width by removing parasitic noise components from the recorded waveforms. This noise suppression improves the performance of the QKD systems under study, particularly in terms of their SKR, by mitigating the effects of instrumental noise. We use the combination of digital filters based on Savitzky-Golay \cite{savitzky1964, bromba1981} and singular value decomposition (SVD) techniques \cite{grassberger1993, konstantinides1997, jha2011, huang2023} (see \cref{sec:data-processing} for details). We calculate the energy of each registered denoised pulse by integrating its area over a fixed time window. Then, from the calculated energy value we determine the pulse's intensity setting. After that, we compute the distributions of pulses' energies for each studied intensity \textit{pattern} ($3^{\xi + 1}$ distributions in total for each analyzed system). As an example, in \cref{fig:experimental-setup}(d), we present the energy distributions for all $\xi=1$ \textit{patterns} together with the overall distributions of pulses' energies for each intensity setting (gray) for system~B. Moreover, we show a zoomed-in sector with decoy-state pulses' energies distributions and their mean values in \cref{fig:experimental-setup}(e). In \Cref{sec:results} we compare these mean values to ascertain whether intensity correlations exist in the systems under study.

\subsection*{Assumptions}
We make the following assumptions during the measurements and data processing.

\textit{Assumption 1.}
We define the intersymbol intensity correlations by observing the recorded energies of bright non-attenuated optical pulses. We assume that the correlations at the single-photon level of energy are the same as those observed at the classical level of optical energy. From our point of view, the optical attenuation is not an active process and should not contribute to intensity correlations, since the energy of each attenuated pulse is reduced equally and independently of each other.

\textit{Assumption 2.}
Since we make our analysis based on experimentally measured values of the optical energy, we assume that our measurement equipment converts the optical power into digital values linearly. Precisely, we consider that the optical-to-electrical conversion in the classical photoreceiver and the voltage measurement in the oscilloscope are linear. To support this assumption, we carefully calibrated the measurement setups to ensure that all recorded signals well fall within the near-linear region of the devices’ dynamic range. Additionally, we selected equipment with sufficient bandwidth to follow the signal slopes linearly. Finally, we conducted all measurements under stable laboratory conditions to minimize external influences. We acknowledge, however, that any residual non-linear effects may influence the estimation of intensity correlations and should be taken into account when establishing a conservative lower bound for system security. While there could still be residual non-linear artifacts, we assume they do not significantly affect our results and we do not evaluate their impact on the security of the tested systems. This assumption is supported by the calibration data and the specifications of the devices used in both QKD system measurements.

\textit{Assumption 3.}
To make sure that instrument noise components do not contribute to the resulting energy calculations, we utilize digital filtering based on the Savitzky-Golay \cite{savitzky1964, bromba1981} and SVD \cite{grassberger1993, konstantinides1997, jha2011, huang2023} techniques. We emphasize that we do not provide formal statistical confidence bounds on the deviation of the filtered waveform from the true signal by using these methods, while this, in principle, is possible \cite{huang2023, oxby2024}. As such, the filtered signals should be regarded as noise-suppressed estimates rather than rigorously bounded quantities. This limitation should be considered when interpreting correlation measurements and assessing the security implications. Also, it is important to note that there is a risk that signal components affected by intensity correlations may be inadvertently filtered out by these methods. Since the true frequency response of intensity fluctuations arising from interpulse correlations is unknown, it is challenging to design a filtering scheme that suppresses only the instrumental noise without also impacting signal components related to these correlations. This presents an additional non-trivial consideration that must be taken into account if the characterization method proposed in this study is to be applied by QKD equipment manufacturers or incorporated into certification procedures, as it may directly influence the lower bound of the system’s security. Nevertheless, in the context of this study, we consider this risk to be negligible. The high-frequency instrumental noise we filter out and the intensity fluctuations caused by correlations are expected to be well-separated in the frequency domain, based on the potential sources of these fluctuations discussed later in the paper. Therefore, we assume that the applied filtering effectively removes the instrumental noise while preserving the integrity of the actual signal.

\textit{Assumption 4.}
The noise floor in the measured data is about $0~\volt$, or even slightly below this value as shown in \cref{fig:experimental-setup}(c). While this is a consequence of the typical unavoidable effect of non-ideal measurement device calibration, thermal noise, electromagnetic interference, and quantization effects due to the finite resolution and sensitivity of the devices, it results in negative values when calculating the energy for V pulses, which obviously has no physical meaning and is a problem for secret key rate calculations. We overcome this issue by adding the lowest negative energy value found within our experimental sequence to each calculated pulse energy. This guarantees that all the pulses in the data set have energy greater or equal to zero after this operation. While this does not affect the experimental results and calculated energy distributions, it can slightly affect the calculation of the secret key rates. We believe that applying a constant shift to ensure positivity represents a pragmatic and minimally invasive solution while keeping the relative differences and correlations between analyzed pulses unchanged. Therefore, we consider it a practical and consistent solution under the current constraints. We assume that the effect of this shift on the secret key rate calculation is small, but a rigorous analysis needs to be done to ensure that this is indeed the case. For that matter, we conducted several  numerical simulations, verifying that the secret key rate is not overestimated and that the intensity shifts described above yield a conservative estimate. However, once again we remark that a rigorous proof confirming the general validity of this assumption lies beyond the scope of the present work.

\section{Results}
\label{sec:results}

\begin{figure*}
\includegraphics{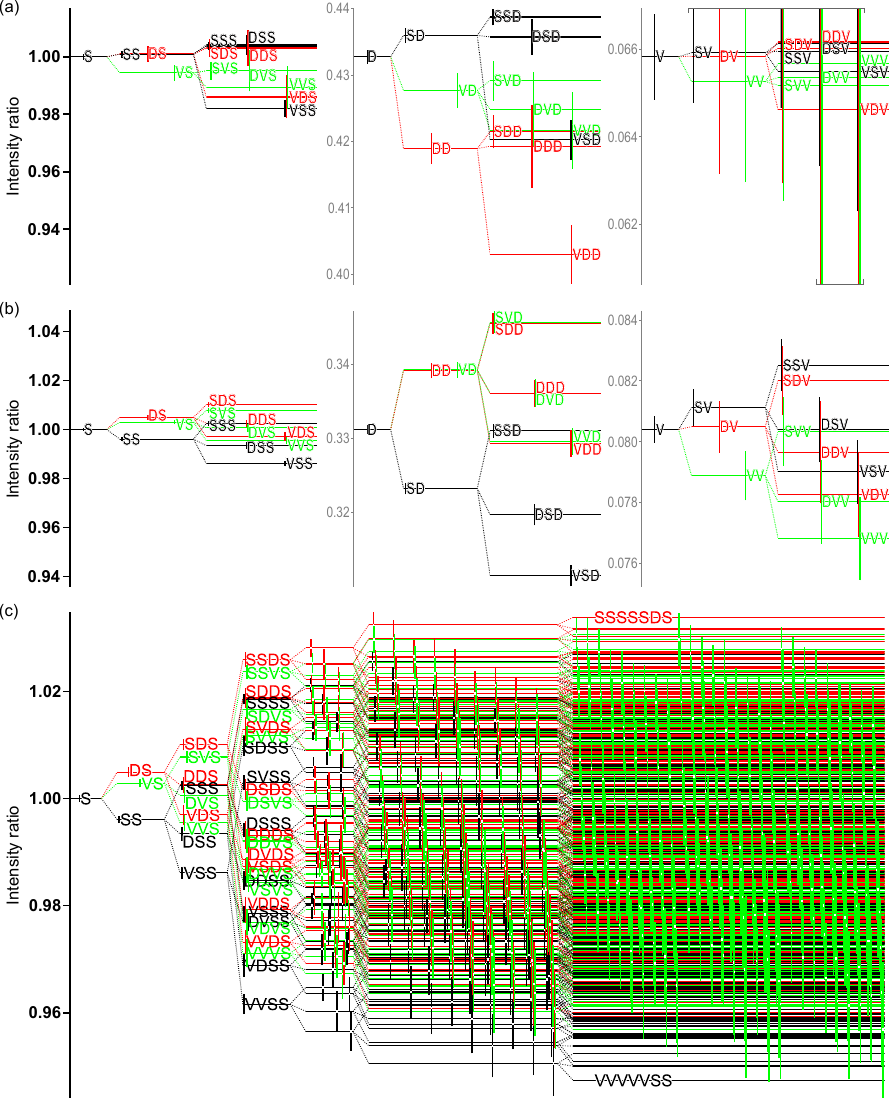} 
\caption{\label{fig:experimental-result} 
Measured intensity correlations in (a) system~A and (b) system~B up to second order ($\xi=2$). The vertical position of each horizontal line represents the mean energy value corresponding to a given \textit{pattern} that is specified by the line's label. The color of the line indicates the setting of the nearest-neighbouring pulse, being black, red or green for S, D or V respectively. The value is normalized to the mean of S (gray vertical scales) or to the mean of the state plotted (black vertical scale, applies to all three sets). The vertical line to the left of each label represents the confidence interval for a confidence level of $0.9$. The experimental results for the S states of system~B are extended up to $\xi=6$ in (c). The most separated \textit{patterns} associated to sixth order correlations are labeled and their waveforms are analyzed in \cref{fig:experimental-result-oscillograms}(a) and (b). Owing to the lack of figure space, in (c) we plot confidence intervals at the sixth order only for those \textit{patterns} whose last two pulses are VS (green). We also provide additional calculated ratios in \cref{sec:data-values}.}
\end{figure*}

\begin{figure}
\includegraphics{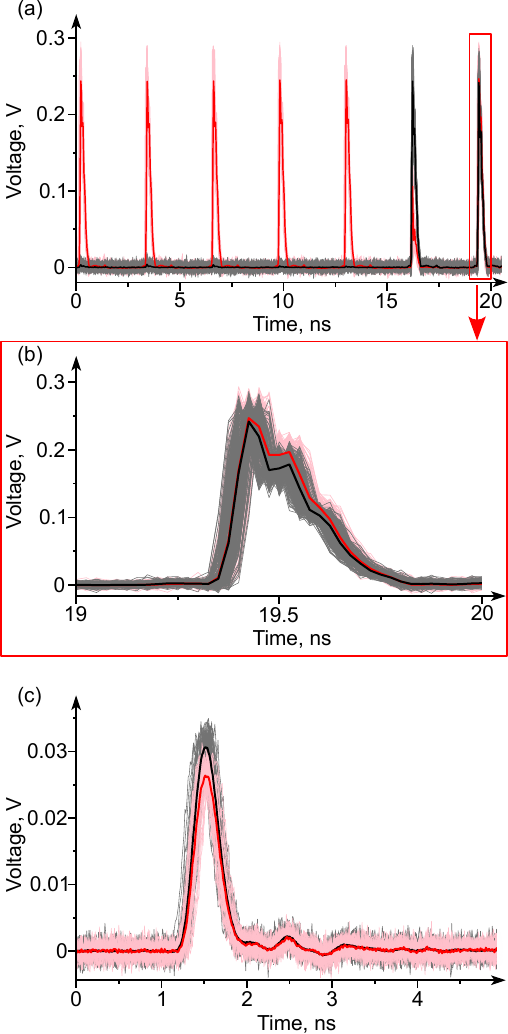} 
\caption{\label{fig:experimental-result-oscillograms}Comparison of averaged oscillogram waveforms of the two most-intensity-separated \textit{patterns}, (a) measured in system~B---SSSSSDS (red solid line) and VVVVVSS (black solid line). Individual measured pulses are plotted in pink (SSSSSDS, 250 pulses) and gray (VVVVVSS, 250 pulses). A noticeable deviation in the waveforms of the S state when considering $\xi=6$ is additionally shown in the zoomed-in sector~(b). A similar effect of the long correlations on the pulses' waveforms is observed in system~A as well~(c). In the latter, we compare the waveforms of the most separated decoy-state third-order \textit{patterns}---DSSD (black and gray, 50 pulses) and VVDD (red and pink, 50 pulses).}
\end{figure}

We analyze a recorded sequence containing 171120 S states, 28240 D states, and 24201 V states produced by system~A (509267~S, 255433~D, and 254139 V states produced by system~B) that we collected during the measurement. We present the central experimental result in \cref{fig:experimental-result}. We show the intensity ratios for the first- and second-order correlations for both systems in \cref{fig:experimental-result}(a) and (b). Here, the vertical position of each horizontal line denotes the mean energy value corresponding to the \textit{pattern} indicated by the label, in relative (black vertical scale) and absolute (gray vertical scales) units. As already mentioned, the rightmost letter of each label indicates the intensity setting of the considered \textit{pattern}. Precisely, the vertical position of every labeled line given by
\begin{subequations}
\begin{align}
\label{vertical-position-1}
\text{position}_\text{rel} &= \frac{\langle \text{pattern} \rangle}{\langle \text{setting} \rangle} \text{ on the black global scale;}\\
\label{vertical-position-2}
\text{position}_\text{abs} &= \frac{\langle \text{pattern} \rangle}{\langle S \rangle} \text{ on the gray local scale.}
\end{align}
\end{subequations}
Here, $\langle \text{pattern} \rangle$ is the mean energy of the rightmost pulse in the label of the analyzed \textit{pattern}, $\langle \text{setting} \rangle$ is the mean energy of the corresponding one-letter distribution, and $\langle S \rangle$ is the mean energy of S for a given QKD system. In \Cref{sec:data-processing} we provide examples illustrating how the intensity ratios are calculated. For each system, there are 39 horizontal lines: 3 of them have one-letter labels and show the mean of the energy distribution of all pulses with the intensity setting S, D, or V; 9 have two letters in the label and represent the mean of the energy distribution for each $\xi=1$ \textit{pattern}; finally, 27 labeled with three letters represent each $\xi=2$ \textit{pattern}. According to \cref{fig:experimental-result}(a) and (b), intensity correlations are present in both studied QKD systems, with setting D being the most affected by them. Furthermore, as can be seen from the same figure, the deviations of the second-order \textit{patterns} for the S and D settings in both systems are either similar or even greater than those corresponding to nearest-neighbour correlations. We examine this long correlation effect even more in \cref{fig:experimental-result}(c), where we illustrate the intensity ratios for the S states of system~B up to $\xi=6$. While it is commonly assumed that the first-order correlations should have the greatest impact on the intensity of the correlated state, our findings suggest that the largest deviations between intensities correspond to the third-order correlated \textit{patterns}. Moreover, as can be seen from the same figure, the strength of the correlations decreases relatively slowly with an increase of the order of correlation length, making a noticeable impact even in the fifth- and sixth-order correlated \textit{patterns}. We note that, for the latter order, we plot the confidence intervals only for the \textit{patterns} whose last two pulses are VS. These are the \textit{patterns} that have less instances in our observed data set and therefore they have the largest confidence intervals. The energy deviations caused by the intensity correlations are almost indistinguishable for these \textit{patterns}, while for the ones whose last two pulses are SS or DS, the sixth-order deviations are still statistically significant. We compare the waveforms of the higher-order labeled \textit{pattern} pulses for both systems in \cref{fig:experimental-result-oscillograms}. The waveforms of the pulses of the same intensity settings clearly tend to have different amplitude and shape.

In the perfect scenario, when a QKD system does not have intensity correlations, all the \textit{patterns} presented in \cref{fig:experimental-result} should form one single horizontal line at the relative intensity ratio of~1. Clearly, this is not the case. Moreover, several lines of the higher-order correlation \textit{patterns}, tend to have a much greater spread (the range between the \textit{patterns} with the highest and lowest energies of the same $\xi$ and ${\langle \text{setting} \rangle}$) than the corresponding nearest-neighbouring distributions. For example, this is observed in the second-order correlations involving the \textit{patterns} of the S and D (D and V) intensity settings of system~A~(B), as well as in the third-order correlations of the S intensity setting \textit{patterns} in system~B. While the deviations in the second-order correlation \textit{patterns} for the S state are less than 2\% in both systems, the second-order \textit{patterns} for the D state exhibit higher deviations (about 7\%) in both systems. In contrast, a notable increase in intensity deviations is observed in the third-order correlation \textit{patterns}; for instance, in system~B, these deviations for the S state \textit{patterns} reach approximately 4\%. This suggests that higher-order correlations are not only present in the tested systems but do not decay rapidly and remain noticeably strong as well. We provide a more detailed analysis of the correlation strengths in \cref{sec:correlation-strengths}.

A possible reason for the existence of such relatively large correlations in the decoy setting \textit{patterns} is an unstable operating point of the intensity modulator while encoding the decoy states \cite{lu2021}. Typically, the operating voltages for the vacuum and signal states are chosen to be at the extremes of the $\cos^{2}$-shaped modulator transfer function, which are quite stable positions. On the other hand, the decoy state modulating voltage is placed at the slope of the transfer function, and any small voltage fluctuation results in relatively high encoded intensity deviations. A potential way to address this vulnerability is through a revised design of the intensity modulator \cite{roberts2018}. For instance, replacing conventional Mach–Zehnder (MZ)-based modulators with dual parallel modulators (DPM) allows modulation of intensity states using only the stable extrema of the transfer function, thereby avoiding the less stable regions along its slope \cite{zhang2020}. Still, this countermeasure should be tested in detail to assess its effectiveness.

Another possible source contributing to higher-order intensity correlations is the residual voltage in the modulators, caused by correlations in the high-frequency electrical signals driving them. These signals are typically generated by control electronics and FPGAs in the source unit, which often have limited bandwidth and imperfect frequency responses, resulting in noticeable correlations in the signal waveforms \cite{yoshino2018, trefilov2021}. A possible countermeasure to mitigate this would be to improve the signal quality with the use of higher-bandwidth electronics and pre-emphasis techniques that can compensate for frequency response imperfections \cite{kang2023}. Reinitializing the modulating electronics before each round of the bit preparation could be also an effective way to suppress correlations provided there is enough time for this between rounds in the QKD system. These measures could significantly enhance the stability and independence of intensity modulation between subsequent pulses.

\section{Effect on the security of QKD}
\label{sec:theory}

\subsection{Theoretical analysis}

To account for the influence of intensity correlations in the decoy-state method, we apply the security analysis presented in~\cite{sixto2022}, based on the so-called Cauchy-Schwarz (CS) constraint \cite{lo2007,pereira2020,zapatero2021}. This result is used to upper-bound the bias that Eve can induce between the detection statistics of Fock states with different records of intensity settings. In what follows, we elaborate on the details of this parameter estimation technique for the case of nearest-neighbour pulse correlations, $\xi=1$, and the reader is referred to~\cite{sixto2022} for further details. Importantly,  this security proof neglects the potential information leakage arising from variations in the shape of the pulses due to intensity correlations (see \cref{fig:experimental-result-oscillograms}).

Firstly, we list the three assumptions on which the analysis relies.

(i) For any given round $k$ and photon number $n_{k}$, there exists a \textit{physical intensity} $\alpha_{k}$ such that
\begin{equation}
\label{poisson}
p\left(n_{k}| \alpha_{k}\right)=\frac{e^{-\alpha_{k}} \alpha_{k}^{n_{k}}}{n_{k} !}.
\end{equation}
Namely, the photon-number statistics are Poissonian conditioned on the value of the physical intensity. This feature is supported by recent high-speed QKD experiments \cite{yoshino2018, grunenfelder2020}.

(ii) $\alpha_{k}$ is a bounded random variable for all $k$ and its distribution $g_{a_{k},a_{k-1}}$ is determined by the present setting, $a_{k}$, and the neighbouring setting, $a_{k-1}$. As a consequence,
\begin{equation}
\begin{aligned}
\label{pns}
&\left.p_{n_{k}}\right|_{a_{k}, a_{k-1}}=\int_{a_{k}^{-}}^{a_{k}^{+}} g_{a_{k},a_{k-1}}(\alpha_{k}) \frac{e^{-\alpha_{k}} \alpha_{k}^{n_{k}}}{n_{k} !} d \alpha_{k},
\end{aligned}
\end{equation}
for all $n_{k}$. Note that, without loss of generality, the boundaries can be expressed as $a_{k}^{\pm}=a_{k}\left(1 \pm \delta^{\pm}_{a_{k}, a_{k-1} }\right)$ for some relative deviations $\delta^{\pm}_{a_{k}, a_{k-1}}$ with respect to $a_{k}$.

(iii) The intensity correlations have a finite range $\xi$. The value of the physical intensity of round $k$, $\alpha_{k}$, is only affected by those previous settings $a_{j}$ with $k-j \leq \xi$. We note that this assumption could be removed by using the recent results in \cite{pereira2024}.

Importantly, the above assumptions enable the desired parameter estimation, summarized in \cref{sec:parameter-estimation}.

\subsection{Asymptotic secret key rate simulations}
Frequently, the post-selection technique~\cite{christandl2009,renner2007,renner2009} is invoked to justify the asymptotic equivalence between the secret key rates with collective and coherent attacks. However, in the presence of pulse correlations, a necessary round-exchangeability property of the post-selection technique is invalidated. In a similar way, correlations invalidate the counterfactual argument often invoked under ideal decoy-state preparation~\cite{zapatero2021}. Hence, a different approach must be followed to define an (as general as possible) asymptotic regime. Particularly, if the variances of the experimental averages vanish asymptotically, the secret key rate attainable against coherent attacks in which long-range interdependencies between detection events are not introduced by Eve's attack (see \cite{zapatero2021} for more details) can be estimated as \cite{zapatero2021, sixto2022}
\begin{equation}
\label{skr}
K_{\infty}=\bar{Z}_{1, \mu, N}^{\mathrm{L}}\left[1-h\left(\frac{\bar{E}_{1, \mu, N}^{\mathrm{U}}}{\bar{X}_{1, \mu, N}^{\mathrm{L}}}\right)\right]-f_{\mathrm{EC}} \bar{Z}_{\mu, N} h\left(E_{\mathrm{tol}}\right),
\end{equation}
for a large enough number of rounds $N$~\cite{zapatero2021}, where $\bar{Z}_{1, \mu, N}^{\mathrm{L}}$ ($\bar{X}_{1, \mu, N}^{\mathrm{L}}$) provides a lower bound on the average number of signal-setting single-photon counts among those events in which both Alice and Bob select the $Z$ ($X$) basis, and $\bar{E}_{1, \mu, N}^{\mathrm{U}}$ provides an upper bound on the average number of signal-setting single-photon error counts among those events in which both users select the $X$ basis. Also, $h(x)$ denotes the binary entropy, $f_{\mathrm{EC}}$ stands for the error correction efficiency, $\bar{Z}_{\mu, N}$ is defined as $\bar{Z}_{\mu, N}={Z}_{\mu, N}/N$ for $Z_{\mu, N}=\sum_{c\in A} Z_{\mu, c, N}$, $Z_{\mu, c, N}$ denoting the number of $Z$ basis counts associated to the record of settings $(a_{k},a_{k-1})=(\mu,c)$ and $E_{\mathrm{tol}}$ denotes the overall error rate observed in the $Z$ basis. The quantities $\bar{Z}_{1, \mu, N}^{\mathrm{L}}$, $\bar{X}_{1, \mu, N}^{\mathrm{L}}$ and $\bar{E}_{1, \mu, N}^{\mathrm{U}}$ can be estimated from the observed gains and error gains via linear programming, as shown in \cite{zapatero2021,sixto2022}.

\begin{figure}
\includegraphics{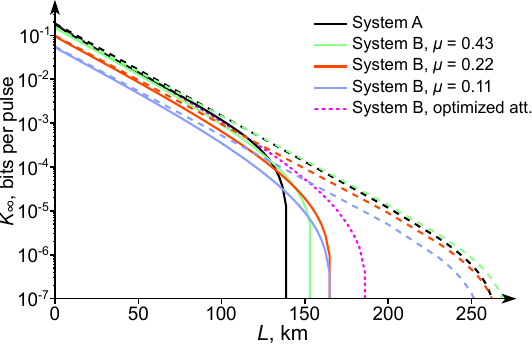} 
\caption{\label{fig:1_vecinos}Asymptotic secret key rate $K_{\infty}$ for the case of first-order pulse correlations ($\xi=1$, solid lines) and for the ideal scenario without correlations ($\xi=0$, dashed lines). For system~B, we consider different attenuations to study the dependence of the secret key rate on the decoy state intensities. It is apparent from the figure that lowering the intensities in the presence of correlations is beneficial for long-distance transmissions. For completeness, we have also included the attainable secret key rate for system~B if the attenuation is optimized for each distance. For the simulations, we use the channel model described in \cref{app:channel}.}
\end{figure}

\begin{figure}
\includegraphics{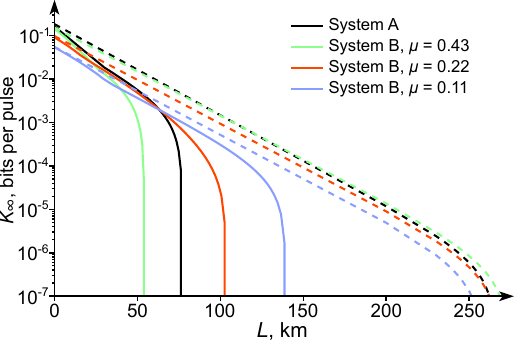} 
\caption{\label{fig:2_vecinos} Asymptotic secret key rate $K_{\infty}$ for the case of second-order pulse correlations ($\xi=2$, solid lines) and for the ideal scenario without correlations ($\xi=0$, dashed lines). Similarly to \cref{fig:1_vecinos}, we consider different values of the attenuation for system~B and we use the channel model presented in \cref{app:channel}.}
\end{figure}

To evaluate the performance of both systems, we assume a truncated Gaussian (TG) distribution for the correlation function $g_{a_{k}, a_{k-1}}$, which is observed in \cref{fig:experimental-setup}(d) and (e) and also motivated by previous studies \cite{yoshino2018, huang2023}. Regardless, the analysis presented below is applicable to any other distribution function. For each system, we set the ratios between the intensity settings, the maximum relative deviations, and the mean and variance of the TG distributions following the experimental values provided in \cref{sec:data-values}. The asymptotic secret key rates calculated is plotted in \cref{fig:1_vecinos,fig:2_vecinos} for $\xi=1$ and $\xi=2$. For system~B, we vary the average intensity of the signal setting $\mu$, which would physically correspond to setting the attenuation of Alice's variable optical attenuator (VOA). The ratios between the intensities and the minimum and maximum deviations are still obtained from \cref{tab:system-b-values}. Note that the signals from system~A could also be further attenuated to improve performance, but since the recorded signals of that system already converted to be at the single-photon level, we opt not to include this additional step.

As can be seen in the plots, increasing the attenuation (i.e.,\ lowering the intensities) is beneficial for long-distance transmission. Besides, it is clear from the figures that considering $\xi=2$ substantially impairs the performance. This is because, as shown in \cite{sixto2022}, there is an exponential dependence with the correlation length in the overlap parameter, which makes the Cauchy-Schwarz constraints looser and causes the different $n$-photon yields associated with different intensity settings to be more distinguishable (see \cref{sec:parameter-estimation} for more details). Due to the exponential growth in the number of constraints with increasing $\xi$, computing the secret key rate for higher-order correlations becomes computationally intensive and often impractical. Therefore, we illustrate our method with $\xi=1,2$ and we further emphasize that the theoretical analysis in \cite{sixto2022} should be viewed as a promising direction rather than a universally applicable solution, as, in general, processing higher-order correlations remains computationally infeasible within this theoretical framework.

\section{Conclusion}
\label{sec:conclusion}

We have experimentally demonstrated the presence of long intensity correlations between the optical pulses produced by two different decoy-state BB84 QKD systems. The impact of higher-order correlations on the pulse's intensity is similar or higher than that of the nearest-neighbour case, even at relatively low sub-gigahertz pulse repetition rates. As discussed in previous literature, this effect challenges a fundamental assumption underlying most decoy-state security proofs, posing a potential threat to the reliability of QKD systems. To address this issue, we have introduced a simple method for measuring the relevant quantities to accurately characterize intensity correlations, and we have assessed their impact on the secret key rate for the first- and second-order correlations, as performing secret key rate simulations for higher-orders becomes computationally challenging with the security analysis employed in this work, based on that in \cite{sixto2022}. Our findings indicate that intensity correlations could substantially impair the performance of decoy-state QKD. Furthermore, we have shown that, according to state-of-the-art security proofs, the secret key rate is highly sensitive to the output mean photon number and ratios between the different intensities. We believe that vendors can optimize these parameters to minimize the effect of intensity correlations on QKD performance. Another strategy could be increasing the bandwidth of the electro-optical devices responsible for the intensity modulation in Alice. Also, we suggest that electrical and optical lines used with these devices should be carefully characterized to avoid parasitic interference, such as multiple back-and-forth reflections in the cables. A preliminary study shows correlations in the electrical signal feeding the modulator \cite{trefilov2021}. From a theoretical standpoint, it seems necessary to develop improved security proofs capable of handling higher-order correlations while remaining computationally scalable.

\acknowledgments
We thank Davide Rusca, Fadri Gr\"{u}nenfelder, Guillermo Curr\'{a}s-Lorenzo, Margarida Pereira, and Ainhoa Agulleiro for discussions.

{\em Funding:} The Ministry of Science and Education of Russia (program NTI center for quantum communications), Russian Science Foundation (grant 21-42-00040), the Galician Regional Government (consolidation of research units: atlanTTic), the Spanish Ministry of Economy and Competitiveness (MINECO), the Fondo Europeo de Desarrollo Regional (FEDER) through the grant No.\ PID2020-118178RB-C21, MICIN with funding from the European Union NextGenerationEU (PRTR-C17.I1), and the Galician Regional Government with own funding through the ``Planes Complementarios de I+D+I con las Comunidades Autonomas'' in Quantum Communication, the ``Hub Nacional de Excelencia en Comunicaciones Cu{\' a}nticas'' funded by the Spanish Ministry for Digital Transformation and the Public Service and the European Union NextGenerationEU, the European Union’s Horizon Europe Framework Programme under the Marie Sk\l{}odowska-Curie Grant No.\ 101072637 (Project QSI), the project “Quantum Security Networks Partnership” (QSNP, grant agreement No.\ 101114043), the National Natural Science Foundation of China (grants 62371459 and 62061136011), and the Innovation Program for Quantum Science and Technology (grant 2021ZD0300704). X.S. acknowledges support from FPI predoctoral scholarship granted by the Spanish Ministry of Science, Innovation and Universities.

{\em Author contributions:} D.T.\ and A.H.\ performed the experiments. D.T.\ analyzed the data. X.S.\ and V.Z.\ performed the simulations to estimate the secure key rate. D.T.\ and X.S.\ wrote the article with help from all authors. V.M.\ and M.C.\ supervised the study.

\section*{Disclosures}
\label{sec:disclosures}

The authors declare no conflicts of interest.

\section*{Data availability}
\label{sec:data-availability}

Data underlying the results presented in this paper are not publicly available at this time but may be obtained from the authors upon reasonable request.

\section*{Methods}
\appendix

\section{Data processing}
\label{sec:data-processing}

To transform the raw oscillogram data collected with the measurement setup [see \cref{fig:experimental-setup}(c)] to the form presented in \cref{fig:experimental-result}, we process the data and apply several filtering techniques to it, namely SVD (for S and D states) and Savitzky-Golay (for V states) digital filters. This is explained below, together with the procedure to calculate the resulting intensity ratios and their confidence intervals shown in \cref{fig:experimental-result}.

\begin{figure}
\includegraphics{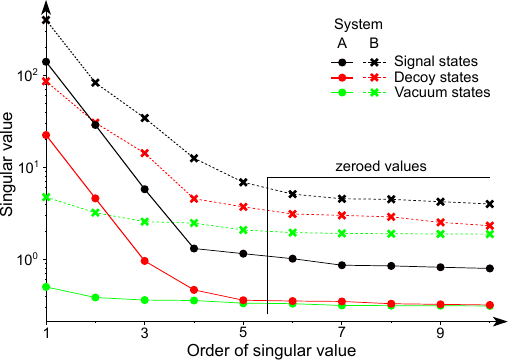}
\caption{\label{fig:singular-values}The first ten orders of singular values of all intensity settings for both examined systems. To remove the noise, we make zero the singular values of the $S$ matrices corresponding to the S and D pulses after the fifth order of singular values.}
\end{figure}

\textit{SVD filtering.}
To apply the SVD filtering method, we populate a matrix $M$ with $n$ recorded oscillograms of noisy pulses of the same intensity setting and width $m$ (the number of points that make up the waveform), so that $M = n \times m$. We can decompose $M$ as
\begin{equation}
\label{eq:SVD_1}
M = USV\tran,
\end{equation}
where $U$ is an $m \times m$ unitary matrix with its columns being left singular vectors, $S$ is an $m \times n$ diagonal matrix with singular values placed in descending order, and $V\tran$ is an $n \times n$ unitary matrix, whose rows are right singular vectors. The initial noisy data in the $M$ matrix after the decomposition is transformed into singular values of the $S$ diagonal matrix. A magnificent property of the SVD method is that the singular values that characterize a waveform of the true measured signal are the first large values of $S$, while the singular values of an independent and individually distributed (i.i.d.)\ Gaussian instrument noise are relatively small and spread along its $m$ dimension. The core idea of the SVD filtering method is that after the decomposition, it is possible to separate the subspaces of true signal singular values from those of the noise, and then one can remove the latter. We show the singular values for all intensity settings in both systems in \cref{fig:singular-values} and indicate the ones that we make zero. As a result, a new diagonal matrix $S'$ can be constructed, with only a few orders of non-zero descending singular values. Then, one can perform the inverse operation of the decomposition to obtain the reconstructed matrix $M'$, populated with filtered optical pulses
\begin{equation}
\label{eq:SVD_2}
M' = US'V\tran.
\end{equation}
After this operation, each row of the resulting matrix $M'$ represents the filtered optical pulse, the noisy version of which was previously contained in the initial matrix $M$. Since the singular values for the V states are relatively low and indistinguishable from those of noise, we cannot filter these optical pulses with this type of filter, so we apply another technique based on the Savitzky-Golay approach, which we present next. 

\begin{figure}
\includegraphics{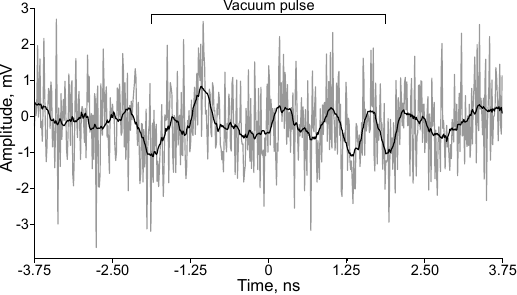}
\caption{\label{fig:vacuums-filtering} An oscillogram of the single vacuum state measured on system~A before (gray) and after (black) Savitzky-Golay filtering.}
\end{figure}

\textit{Savitzky-Golay filtering.}
To filter the recorded vacuum states from the instrument noise, we employ a technique based on a well-known Savitzky-Golay digital low-pass filter with predefined optimized parameters \cite{savitzky1964, bromba1981}. This method is based on a local least-squares low-degree polynomial approximation, and, similarly to the moving-average filter, it smooths the noisy oscillograms by locally fitted polynomial functions at every point of the experimental data. As a result, the distorted signal is smoothed, while the shape of recorded waveforms is maintained. We show an example of the vacuum state before and after filtering in \cref{fig:vacuums-filtering}. For both systems, we optimize and put the degree of fitting polynomial function as 3, and the width of the filtration window as 39 experimental points. 

\begin{figure*}
\includegraphics{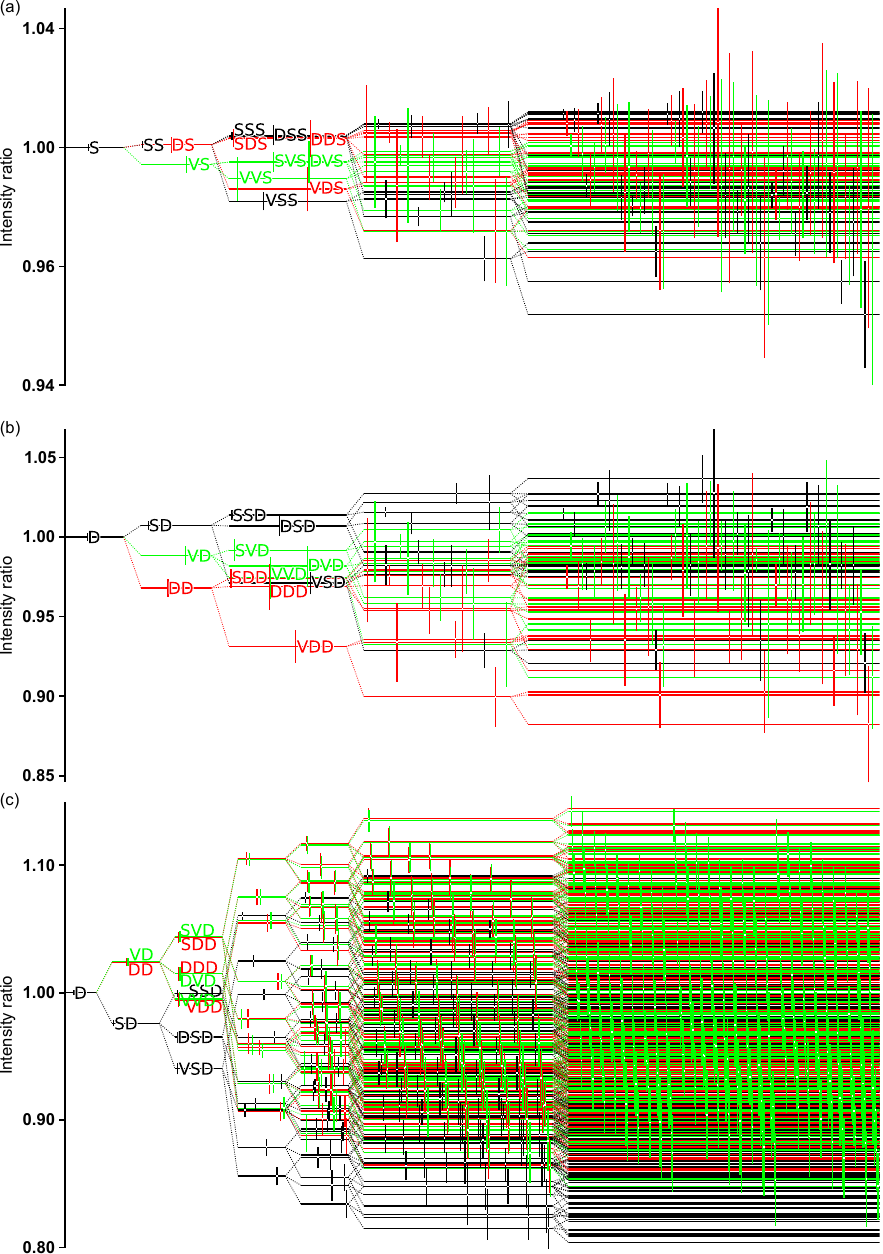}
\caption{\label{fig:experimental-result-appendix}Measured intensity correlations of (a)~the S and (b)~the D states of system~A, and (c)~the D states of system~B. The vertical lines represent the confidence intervals for a confidence level of $0.9$. The lines' colors correspond to those used in \cref{fig:experimental-result}.}
\end{figure*}

\textit{Calculation of confidence intervals.} For clarity, in what follows we provide the confidence intervals for the case of nearest-neighbour correlations, and the generalization to higher-order correlations is straightforward. Let us consider the set of all rounds $k$ such that: (i) setting $a$ is selected in round $k$ and (ii) setting $c$ is selected in round $k-1$. We shall assume that the intensities measured in any two rounds of this set are independent random variables. Under this assumption, one can infer a confidence interval on the population mean of the set $\langle \alpha_{ac} \rangle$, given the sample mean $\bar{\alpha}_{ac}$, using Hoeffding's inequality. Particularly, for a confidence level of $1-\delta$, the interval reads
\begin{equation}
\label{confidence-intervals-1}
I_{ac}=[\bar{\alpha}_{ac}-\Delta_{\delta/2}, \bar{\alpha}_{ac}+\Delta_{\delta/2}],
\end{equation}
where $\Delta_{\delta/2}=w_{ac}\sqrt{\ln(2/\delta)/2N_{ac}}$, $N_{ac}$ is the number of \textit{patterns} $ac$ in the data set, and $w_{ac}$ denotes the maximum range among all random variables in the set. Since the theoretical ranges are unknown, to estimate $I_{ac}$ we replace $w_{ac}$ with the difference between the largest and the smallest measured intensities in the set. For the confidence intervals that we plot in \cref{fig:experimental-result,fig:experimental-result-appendix} we fix the value of $\delta = 0.1$. As can be seen from those figures, the intervals get wider with the increase of the correlation length, which is a direct consequence of the fact that $I_{ac}$ is inversely proportional to the number of accumulated counts $N_{ac}$.

\textit{Examples of calculation.} Here, we briefly explain how we calculate the vertical positions of the lines plotted in \cref{fig:experimental-result} for D, VS, and SDV \textit{patterns} of the system~A. According to \cref{vertical-position-1,vertical-position-2},
\begin{equation}
\label{vertical-position-ex-1}
\begin{split}
D_\text{rel} & = \frac{\langle D \rangle}{\langle D \rangle} = 1,\\
VS_\text{rel} & = \frac{\langle VS \rangle}{\langle S \rangle} = \frac{0.635059}{0.638635} \approx 0.9944,\\
SDV_\text{rel} & = \frac{\langle SDV \rangle}{\langle V \rangle} = \frac{0.042150}{0.042044} \approx 1.0025,
\end{split}
\end{equation}
\begin{equation}
\label{vertical-position-ex-2}
\begin{split}
D_\text{abs} & = \frac{\langle D \rangle}{\langle S \rangle} = \frac{0.276372}{0.638635} \approx 0.4328,\\
VS_\text{abs} & = \frac{\langle VS \rangle}{\langle S \rangle} = \frac{0.635059}{0.638635} \approx 0.9944,\\
SDV_\text{abs} & = \frac{\langle SDV \rangle}{\langle S \rangle} = \frac{0.042150}{0.638635} \approx 0.06600.
\end{split}
\end{equation}
The values we use in these equations are average energies for distributions of corresponding \textit{patterns} that we converted into photons for system~A (arbitrary units for system~B). We provide this data for all \textit{patterns} and both QKD systems in \cref{sec:data-values}.

\section{Parameter estimation technique}
\label{sec:parameter-estimation}

In this section, we present the derivation of the necessary constraint to estimate the relevant parameters via linear programming \cite{zapatero2021, sixto2022}.

For any given round $k$ and photon-number $n$, the yield and the error probability associated to the pair of settings $(a_{k},a_{k-1})=(a,c)$ are defined as
\begin{equation}
\label{yield}
\begin{aligned}
&Y_{n, a,c}^{(k)}=p(s_{k} \neq f|\\
&n_{k}=n, a_{k}=a, a_{k-1}=c, x_{k}=Z, y_{k}=Z),\\
&H_{n, a,c,r}^{(k)}=p(s_{k} \neq f, s_{k} \neq r_{k}|\\
&n_{k}=n, a_{k}=a,a_{k-1}=c, x_{k}=X, y_{k}=X, r_{k}=r),
\end{aligned}
\end{equation}
where $x_{k} \in\{X, Z\}$, $r_{k} \in \{0,1\}$ represent the key bits selected by Alice, $y_{k} \in\{X, Z\}$ represents Bob's basis selection, $s_k$ stands for Bob's classical outcome, and $f$ represents a ``no-click'' event.

In virtue of the CS constraint, for any pair of distinct settings $a$ and $b$ in round $k$, and for any setting $c$ in round $k-1$, the associated yields and error probabilities satisfy
\begin{equation}
\label{cs_text}
G_{-}\!\left(Y_{n, a,c}^{(k)}, \tau_{a, b,c, n}\right) \leq Y_{n, b,c}^{(k)} \leq G_{+}\!\left(Y_{n, a,c}^{(k)}, \tau_{a, b,c, n}\right)
\end{equation}
and
\begin{equation}
\label{cs_text2}
G_{-}\!\left(H_{n, a,c, r}^{(k)}, \tau_{a, b,c, n}\right) \leq H_{n, b,c, r}^{(k)} \leq G_{+}\!\left(H_{n, a,c, r}^{(k)}, \tau_{a, b,c, n}\right)\!,
\end{equation}
where
\begin{equation}
\label{eq6}
\begin{aligned}
&G_{-}(y, z)= \begin{cases} g_{-}(y, z) & \text { if } y>1-z \\
0 & \text { otherwise }\end{cases},\\
&G_{+}(y, z)= \begin{cases}g_{+}(y, z) & \text { if } y<z \\
1 & \text { otherwise }\end{cases},
\end{aligned}
\end{equation}
and the functions $g_{\pm}(y, z)=y+(1-z)(1-2 y) \pm 2 \sqrt{z}(1-z) y(1-y)$. That is, \cref{cs_text,cs_text2} quantify how much $Y_{n, b,c}^{(k)}$ and $H_{n, b,c, r}^{(k)}$ can deviate from $Y_{n, a,c}^{(k)}$ and $H_{n, a,c, r}^{(k)}$, respectively. Crucially, $\tau_{a, b,c, n}$ is the overlap parameter which quantifies the tightness of the constraints. It represents a lower bound on the squared overlap between the two quantum states underlying the two yields that enter the constraints (see~\cite{sixto2022} for further details).

The fact that $g_{a_{k}a_{k-1}}$ follows a truncated Gaussian distribution, allows to derive explicit formulas for the overlap parameter in \cref{cs_text,cs_text2}, and the photon-number statistics of \cref{pns}. As shown in \cite{sixto2022}, the overlap can be computed with the following formula
\begin{equation}
\tau_{a,b,c,n}^{\xi=1}=\left[\sum_{a_{k+1}\in A} p_{a_{k+1}}\sum_{n_{k+1}=0}^{n_{\text{cut}}}\sqrt{p_{n_{k+1}}|_{a_{k+1},a}p_{n_{k+1}}|_{a_{k+1},b}}\right]^{2} ,
\end{equation}
where $a_{k+1}$ is the setting selected in round $k+1$ and $p_{a_{k+1}}$ represents the probability of selecting such setting. Importantly, for higher order correlations the previous equation presents an exponential behavior with $\xi$ as shown in \cite{sixto2022}, which makes the CS constraints looser and deteriorates the performance. Moreover, the CS constraints can be linearized so as not to break the linear character of the parameter estimation. Importantly, when applying the linearization step to the constraints in \cref{cs_text} [\cref{cs_text2}], an additional reference yield (error) parameter needs to be incorporated (see \cite{zapatero2021, sixto2022} for more details). In these works, the authors do not optimize these parameters to maximize the key rate, as they take the reference yield (reference error) as the one that can be expected by the behavior of the channel, neglecting the effect of correlations. This leads to a severe drop in performance, especially when the outputs of the linear programs and the reference parameters are mutually distant. To fix this and improve the secret key rate, we use the outputs of the linear programs at a certain distance point $L$ as the reference values for the next distance point $L+1$. For the first point, $L=0$, we find a close-to-optimal reference values by running the linear program multiple times with different reference parameters, and using the highest output, in Monte Carlo fashion. Note that the inflection points observed in the secret key rates of \cref{fig:2_vecinos} stem from the difference in how the first point is optimized compared to the rest. Ideally, we would perform a Monte Carlo simulation for all distances, but this is impractical given the slow performance of the linear programs described in \cite{sixto2022}. Importantly, the linear constraints represent a valid bound regardless of the reference value used.

As for the decoy-state constraints, their derivation is rather standard, and they can be written as
\begin{equation}
\begin{aligned}
&\frac{\left\langle\bar{Z}_{a, c, N}\right\rangle}{q_{\mathrm{Z}}^{2} p_{a} p_{c}}\geq\sum_{n=0}^{n_{cut}} p_{n|a,c} y_{n, a, c, N},\\
&\frac{\left\langle\bar{Z}_{a, c, N}\right\rangle}{q_{\mathrm{Z}}^{2} p_{a} p_{c}} \leq 1- \sum_{n=0}^{n_{cut}} p_{n|a,c} + \sum_{n=0}^{n_{cut}} p_{n|a,c} y_{n, a, c, N},
\end{aligned}
\end{equation}
and
\begin{equation}
\begin{aligned}
&\frac{\left\langle\bar{E}_{a, c, N}\right\rangle}{q_{\mathrm{X}}^{2} p_{a} p_{c}}\geq\sum_{n=0}^{n_{cut}} p_{n|a,c} h_{n, a, c, N},\\
&\frac{\left\langle\bar{E}_{a, c, N}\right\rangle}{q_{\mathrm{X}}^{2} p_{a} p_{c}} \leq 1- \sum_{n=0}^{n_{cut}} p_{n|a,c} + \sum_{n=0}^{n_{cut}} p_{n|a,c} h_{n, a, c, N}.
\end{aligned}
\end{equation}
Here, $q_{Z(X)}$ denotes the probability of selecting the $Z$ ($X$) basis, $p_{a}$ denotes the probability of selecting setting $a\in{}A$ in any given round, and the average yields and error probabilities are defined as $y_{n, a, c, N}=\sum_{k=1}^{N} Y_{n, a, c}^{(k)}/{N}$ and $h_{n, a, c, N}=\sum_{k=1}^{N} H_{n, a, c}^{(k)}/{N}$ for all possible settings. Also, we recall that the average gains are $\bar{Z}_{a, c, N}=Z_{a, c, N}/{N}$ for all possible inputs, and we have defined the average error gains as $\bar{E}_{a, c, N}=E_{a, c, N}/{N}$ where $E_{a, c, N}$ represents the error gain associated to settings $(a,c)$ and the number of rounds $N$.
 
Complementing these constraints with the ones in \cref{cs_text,cs_text2} we can readily bound the key-rate parameters $\bar{Z}_{1, \mu, N}^{\mathrm{L}}$, $\bar{X}_{1, \mu, N}^{\mathrm{L}}$ and $\bar{E}_{1, \mu, N}^{\mathrm{U}}$ via linear programming. As an example, $\bar{Z}_{1, \mu, N}^{\mathrm{L}}$ is computed minimizing the average number of signal-setting single-photon counts among those events where Alice and Bob select the $Z$ basis, which is given by
\begin{equation}
\langle \bar{Z}_{1, \mu, N}\rangle
=\sum_{h\in A}q_{\mathrm{Z}}^{2} p_{\mu}p_{h}p^{(k)}(1|\mu, h) y_{1, \mu, h, N}
\end{equation}
restricted to the above constraints.

\section{Channel model}\label{app:channel}

Let $\eta_{\mathrm{det}}$ denote the common detection efficiency of Bob's detectors, and let $\eta_{\mathrm{ch}}=10^{-\alpha_{\text{att}} L / 10}$ be the transmittance of the quantum channel, where $\alpha_{\mathrm{att}}$ represents the attenuation coefficient of the fiber and $L$ (km) is the distance. Also, let $p_{\mathrm{d}}$ denote the dark count probability of each of Bob's detectors and let $\delta_{A}$ stand for the polarization misalignment occurring in the channel. For the simulations presented in \cref{fig:1_vecinos,fig:2_vecinos}, we use the values $\eta_{\mathrm{det}} = 0.65$, $\alpha_{\mathrm{att}}= 0.2~\deci\bel\per\kilo\meter$, $p_{\mathrm{d}} = 7.2 \times 10^{-8}$, and $\delta_{A} = 0.08$, which are taken from \cite{yin2016}.

As shown in \cite{sixto2022}, a standard model yields
\begin{equation}
\frac{\left\langle\bar{Z}_{a, c, N}\right\rangle}{q_{\mathrm{Z}}^{2} p_{a} p_{c}}=\frac{\left\langle\bar{X}_{a, c, N}\right\rangle}{q_{\mathrm{X}}^{2} p_{a} p_{c}}=1-\left(1-p_{\mathrm{d}}\right)^{2} e^{-\eta a} 
\end{equation}
and
\begin{eqnarray}
\frac{\left\langle\bar{E}_{a, c, N}\right\rangle}{q_{X}^{2} p_{a} p_{c}}&=&\frac{\left\langle\bar{E}_{a, c, N(\mathrm{Z})}\right\rangle}{q_{Z}^{2} p_{a} p_{c}}=\frac{p_{\mathrm{d}}^{2}}{2}+p_{\mathrm{d}}\left(1-p_{\mathrm{d}}\right)\nonumber\\
&\times&\left(1+h_{\eta, a, c, \delta_{\mathrm{A}}}\right)+\left(1-p_{\mathrm{d}}\right)^{2}\nonumber\\
&\times&\left(\frac{1}{2}+h_{\eta, a, c, \delta_{\mathrm{A}}}-\frac{1}{2} e^{-\eta a}\right),
\end{eqnarray}
where $\eta=\eta_{\mathrm{det}} \eta_{\mathrm{ch}}$ represents the total attenuation and $a,c\in A$. The parameter $h_{\eta, a, c, \delta_{\mathrm{A}}}$ is defined as $h_{\eta, a, c, \delta_{\mathrm{A}}}=\frac{1}{2}(e^{-\eta a \cos ^{2} \delta_{\mathrm{A}}}-e^{-\eta a \sin ^{2} \delta_{\mathrm{A}}})$ and $\bar{E}_{a, c, N(\mathrm{Z})}$ is equivalent to $\bar{E}_{a, c, N}$ but referred to the $Z$-basis error clicks instead. The tolerated bit error rate of the sifted key is set to $E_{\mathrm{tol}}=\left\langle\bar{E}_{\mu, N(\mathrm{Z})}\right\rangle /\left\langle\bar{Z}_{\mu, N}\right\rangle$.

\section{Experimental data values}
\label{sec:data-values} 

\renewcommand{\arraystretch}{0.8}
\begin{table}
	\vspace{-0.7em}
  \caption{Experimental values for system~A. All the values are converted into photons.}
		\footnotesize
    \begin{tabular}[t]{ll@{\quad}l@{\quad}ll}
    \hline\hline
    \textit{pattern} & \multicolumn{1}{c}{$\mu$}\quad & \multicolumn{1}{c}{$\sigma$}\quad & min energy & max energy \\
    \hline
    S     & 0.638635 & 0.025245 & 0.5286737 & 0.75879511 \\
    D     & 0.276372 & 0.011834 & 0.2332107 & 0.32248299 \\
    V     & 0.042044 & 0.011438 & 0     & 0.0870799 \\
    SS    & 0.639209 & 0.02523 & 0.5341994 & 0.75879511 \\
    SD    & 0.278423 & 0.011295 & 0.2332107 & 0.32248299 \\
    SV    & 0.042104 & 0.011465 & 0     & 0.0870799 \\
    DS    & 0.639251 & 0.025164 & 0.5332007 & 0.73899578 \\
    DD    & 0.267539 & 0.011064 & 0.2353542 & 0.3072787 \\
    DV    & 0.042053 & 0.0114 & 0.0003578 & 0.08378114 \\
    VS    & 0.635059 & 0.024665 & 0.5286737 & 0.74574953 \\
    VD    & 0.273149 & 0.010465 & 0.2373067 & 0.31787876 \\
    VV    & 0.041681 & 0.011308 & 0.0095704 & 0.08159895 \\
    SSS   & 0.641041 & 0.02466 & 0.5341994 & 0.75879511 \\
    SSD   & 0.280194 & 0.010597 & 0.2394827 & 0.32248299 \\
    SSV   & 0.042127 & 0.011459 & 0     & 0.08584321 \\
    SDS   & 0.640664 & 0.024714 & 0.5332007 & 0.73899578 \\
    SDD   & 0.269188 & 0.010417 & 0.2358168 & 0.30376055 \\
    SDV   & 0.042168 & 0.011334 & 0.0003578 & 0.08238706 \\
    SVS   & 0.635516 & 0.02442 & 0.5286737 & 0.74086963 \\
    SVD   & 0.274081 & 0.010288 & 0.2373067 & 0.31787876 \\
    SVV   & 0.041628 & 0.011378 & 0.0095704 & 0.08159895 \\
    DSS   & 0.641269 & 0.024331 & 0.5344015 & 0.75844048 \\
    DSD   & 0.278274 & 0.010504 & 0.2431097 & 0.31491301 \\
    DSV   & 0.042238 & 0.011337 & 0.0052164 & 0.08045127 \\
    DDS   & 0.640541 & 0.024997 & 0.5598773 & 0.72400049 \\
    DDD   & 0.26778 & 0.01083 & 0.24119 & 0.3072787 \\
    DDV   & 0.04226 & 0.011751 & 0.0058803 & 0.08378114 \\
    DVS   & 0.635509 & 0.024512 & 0.5545599 & 0.74574953 \\
    DVD   & 0.271324 & 0.010368 & 0.2466763 & 0.30586557 \\
    DVV   & 0.04174 & 0.011159 & 0.0112804 & 0.07264567 \\
    VSS   & 0.627076 & 0.02533 & 0.5379419 & 0.73671844 \\
    VSD   & 0.268378 & 0.010408 & 0.2332107 & 0.31405785 \\
    VSV   & 0.04183 & 0.011623 & 0.0016062 & 0.0870799 \\
    VDS   & 0.629704 & 0.025812 & 0.5388653 & 0.7387814 \\
    VDD   & 0.257378 & 0.009283 & 0.2353542 & 0.28204715 \\
    VDV   & 0.041267 & 0.011361 & 0.0054174 & 0.07010183 \\
    VVS   & 0.631902 & 0.025994 & 0.5338721 & 0.73724771 \\
    VVD   & 0.269303 & 0.010501 & 0.2380967 & 0.3014277 \\
    VVV   & 0.041942 & 0.011024 & 0.011866 & 0.07716827 \\
    \hline\hline
    \end{tabular}
  \label{tab:system-a-values}
\end{table}

\begin{table}
	\vspace{-0.7em}
  \caption{Experimental values in arbitrary units for system~B normalized to the mean value of S.}
		\footnotesize
    \begin{tabular}[t]{ll@{\quad}l@{\quad}ll}
    \hline\hline
    \textit{pattern} & \multicolumn{1}{c}{$\mu$}\quad & \multicolumn{1}{c}{$\sigma$}\quad & min energy & max energy \\
    \hline
    S     & 1     & 0.03082 & 0.83158619 & 1.16027652 \\
    D     & 0.331205 & 0.031052 & 0.19323125 & 0.45560021 \\
    V     & 0.080408 & 0.019021 & 0     & 0.17750194 \\
    SS    & 0.996082 & 0.030962 & 0.83158619 & 1.11589638 \\
    SD    & 0.32323 & 0.030018 & 0.19323125 & 0.42464045 \\
    SV    & 0.081123 & 0.01879 & 0.00200947 & 0.17750194 \\
    DS    & 1.004939 & 0.030197 & 0.88605649 & 1.16027652 \\
    DD    & 0.339161 & 0.030004 & 0.24043238 & 0.44989812 \\
    DV    & 0.080498 & 0.019254 & 0     & 0.17635141 \\
    VS    & 1.002882 & 0.030115 & 0.87277413 & 1.11230916 \\
    VD    & 0.339283 & 0.029979 & 0.23637931 & 0.45560021 \\
    VV    & 0.078888 & 0.019157 & 0.00130488 & 0.1677094 \\
    SSS   & 1.002422 & 0.030046 & 0.86989206 & 1.11589638 \\
    SSD   & 0.331017 & 0.029067 & 0.23296353 & 0.42464045 \\
    SSV   & 0.082523 & 0.018923 & 0.00378569 & 0.17750194 \\
    SDS   & 1.010071 & 0.029604 & 0.89394137 & 1.1189707 \\
    SDD   & 0.34559 & 0.029354 & 0.24230907 & 0.44989812 \\
    SDV   & 0.082017 & 0.019482 & 0     & 0.16690062 \\
    SVS   & 1.007758 & 0.029584 & 0.89158475 & 1.11210491 \\
    SVD   & 0.345728 & 0.029272 & 0.24372433 & 0.45560021 \\
    SVV   & 0.080346 & 0.019318 & 0.00130488 & 0.1677094 \\
    DSS   & 0.993388 & 0.030198 & 0.83158619 & 1.09894054 \\
    DSD   & 0.319657 & 0.028884 & 0.20888549 & 0.42075485 \\
    DSV   & 0.080396 & 0.01868 & 0.00217222 & 0.15755742 \\
    DDS   & 1.002537 & 0.029783 & 0.88605649 & 1.10954641 \\
    DDD   & 0.336109 & 0.029079 & 0.24827166 & 0.43495529 \\
    DDV   & 0.079659 & 0.018802 & 0.00307564 & 0.17635141 \\
    DVS   & 1.000633 & 0.029598 & 0.88648747 & 1.11230916 \\
    DVD   & 0.336097 & 0.029045 & 0.23993033 & 0.43421851 \\
    DVV   & 0.07803 & 0.018938 & 0.01017669 & 0.15310763 \\
    VSS   & 0.986114 & 0.030495 & 0.86121691 & 1.10345797 \\
    VSD   & 0.311454 & 0.028425 & 0.19323125 & 0.41074145 \\
    VSV   & 0.079018 & 0.018386 & 0.00200947 & 0.15432216 \\
    VDS   & 0.996976 & 0.02977 & 0.88986999 & 1.16027652 \\
    VDD   & 0.329286 & 0.029034 & 0.24043238 & 0.42393855 \\
    VDV   & 0.078274 & 0.018972 & 0.00509359 & 0.14976811 \\
    VVS   & 0.995446 & 0.029884 & 0.87277413 & 1.1006765 \\
    VVD   & 0.329591 & 0.029173 & 0.23637931 & 0.42357123 \\
    VVV   & 0.076824 & 0.018809 & 0.01018254 & 0.15123735 \\
    \hline\hline
    \end{tabular}
  \label{tab:system-b-values}
\end{table}

We provide the experimental values that characterize the studied intensity \textit{patterns'} distributions for system~A (system~B) in \cref{tab:system-a-values} (\cref{tab:system-b-values}). The represented parameters are the mean energy value $\mu$, standard deviation $\sigma$, and minimum and maximum energy values in a given distribution. All the described parameters are converted to be in photons (arbitrary units) for system~A (system~B) and correspond to the truncated Gaussian distributions. We also plot intensity ratios for the S and D states of system~A and the D state of system~B in \cref{fig:experimental-result-appendix}.

With these values it is possible to reproduce the key rate figures in \cref{fig:1_vecinos,fig:2_vecinos}. However, for system~B, we suppose that  a VOA is placed at the output of Alice's source. This corresponds to scaling all the intensities in \cref{tab:system-b-values} by a certain factor. For instance, to reproduce the simulations with $\mu=0.22$, all intensity \textit{patterns} should be scaled by this factor of $0.22$.

\begin{figure*}
\includegraphics{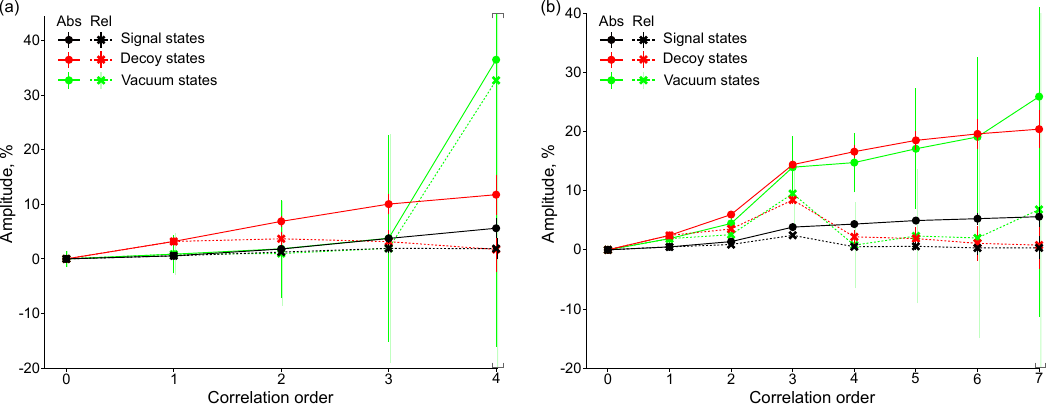}
\caption{\label{fig:correlation-strengths} 
Calculated intensity correlation strengths for each analyzed $\xi$ value. These are defined as the differences between the most deviated \textit{pattern} from the mean value of the corresponding setting energy distribution in (a) system~A and (b) system~B. The solid lines represent how the $\text{Strength}_{\text{Abs}}(\xi)$ parameter varies with the correlation order, while the dashed lines show the corresponding changes in $\text{Strength}_{\text{Rel}}(\xi)$. The color of the lines reveals the intensity setting affected by the correlations, and the vertical lines represent the confidence intervals for a confidence level of 0.9. In both (a) and (b), changes in $\text{Strength}_{\text{Rel}}(\xi)$ show that intensity correlations are decaying with the higher orders. At the same time, in certain cases (e.g.,\ $\xi = 2$[$\xi = 3$] for system~A[B]) higher-order correlations can be significantly stronger than those observed in the nearest-neighbour case.}
\end{figure*}

\section{Correlation strength definition}
\label{sec:correlation-strengths} 

While we provide the general experimental results in \cref{fig:experimental-result,fig:experimental-result-appendix}, it might be quite difficult to directly evaluate the actual behavior of intensity correlations from these figures, primarily showing the intensity ratios for the tested \textit{patterns}. To ease this task, in \cref{fig:correlation-strengths}, for both tested QKD systems we provide the strengths of the intensity correlations for each analyzed $\xi$ value as the difference between the \textit{patterns} most affected by intensity correlations and their corresponding $\langle \text{setting} \rangle$ parameters. We calculate these differences as follows:

\begin{subequations}
\begin{align}
\label{strength-abs}
\text{Strength}_{\text{Abs}}(\xi) &= \frac{\max\left( \left| \langle \text{pattern} \rangle_{\xi} - \langle \text{setting} \rangle \right| \right)}{\langle \text{setting} \rangle},\\
\label{strength-rel}
\text{Strength}_{\text{Rel}}(\xi) &= \text{Strength}_{\text{Abs}}(\xi) - \text{Strength}_{\text{Abs}}(\xi - 1),
\end{align}
\end{subequations}
where, similarly to Eqs.~(1a) and (1b), $\langle \text{pattern} \rangle_{\xi}$ is the mean energy of the considered \textit{pattern} distribution at a given $\xi$, and $\langle \text{setting} \rangle$ the mean energy of the corresponding one-letter distribution. The parameter $\text{Strength}_{\text{Rel}}(\xi)$, in turn, represents the ``gain'' in the strength of the correlations, reflecting their dynamics, while $\text{Strength}_{\text{Abs}}(\xi)$ is the difference between the most-deviated \textit{pattern} and its $\langle \text{setting} \rangle$.

\Cref{fig:correlation-strengths} shows that decoy-setting \textit{patterns} are most strongly affected by intensity correlations, with the largest deviations in the mean values of the \textit{patterns}' energy distributions observed in both systems being more than 10\%~(20\%) for system~A(B) for the highest $\xi$ examined. Additionally, \cref{fig:correlation-strengths}(b) clearly indicates that the strongest correlation order occurs at $\xi = 3$, where the largest relative deviations in the mean values of the \textit{patterns}' energy distributions are observed, based on the $\text{Strength}_{\text{Rel}}(\xi)$ parameter for all three intensity settings. Furthermore, the analysis of $\text{Strength}_{\text{Rel}}(\xi)$ reveals that the relative strength of the correlations gradually decreases as the higher $\xi$ orders are approached, except for the vacuum setting \textit{patterns}, where conclusions are difficult due to their large confidence intervals. This suggests that, as expected, the correlations weaken at higher orders.

\def\bibsection{\medskip\begin{center}\rule{0.5\columnwidth}{.8pt}\end{center}\medskip} 

\bibliography{library}
\end{document}